\documentclass[aps,prl,twocolumn,superscriptaddress]{revtex4-1}

\usepackage{url}
\usepackage{footmisc}
\usepackage{graphicx}
\begin{document}

%\preprint{}

\title{Complexity, Centralization, and Fragility in Economic Networks}


\author{Carlo Piccardi}
\email[Corresponding author: ]{carlo.piccardi@polimi.it}
%\thanks{}
\affiliation{Department of Electronics, Information, and Bioengineering, Politecnico di Milano, Piazza Leonardo da Vinci 32, 20133 Milano, Italy}
\author{Lucia Tajoli}
%\email[]{carlo.piccardi@polimi.it}
%\thanks{}
\affiliation{Department of Management, Economics, and Industrial Engineering, Politecnico di Milano, Via Lambruschini 4B, 20156 Milano, Italy}


\date{\today}

\begin{abstract}
Trade networks, across which countries distribute their products, are crucial components of the globalized world economy. Their structure is strongly heterogeneous across products, given the different features of the countries which buy and sell goods. By using a diversified pool of indicators from network science and product complexity theory, we quantitatively confirm the intuition that, overall, products with higher complexity -- i.e., with larger technological content and number of components -- are traded through a more centralized network -- i.e., with a small number of countries concentrating most of the export flow. Since centralized networks are known to be more vulnerable, we argue that the current composition of production and trading is associated to high fragility at the level of the most complex -- thus strategic -- products.
\end{abstract}

%\keywords{}

\maketitle

%\section{}
%\subsection{}
%\subsubsection{}

%%%%%%%%%%%%%%%%%%%%%

How fragile is the world economy? Given the increasing globalization of economic systems, will economic shocks have widespread diffusion to all countries? The recent evidence of the international financial crisis of 2007-2008 and the European debt crisis suggest that indeed most of the world countries are highly exposed. According to the International Monetary Fund \cite{imf:16}, very few countries were spared the trade slowdown that followed the crisis. Also, non-economic events hitting a specific economy have a broad impact: the volcano eruption in Iceland in 2010 and the earthquake in Japan in 2011 generated significant effects on production, not only in the countries directly hit, but also in a number of other economically linked economies \cite{BeMu:14,CaNi:16}.  These events stirred a debate on the relevance and persistence of transmission of such economic shocks, which however has not reached a definitive conclusion. In fact, while the increased diversification of economic links between countries \cite{DeTa:11,DeTa:14} should make them more resilient, the high density of the world trade network helps the rapid diffusion of shocks \cite{ScFa:09,Ve:10}.

This work contributes to this debate by showing that the high density of the economic links among countries occurs together with a very uneven distribution of such links. Using different indicators, we consistently observe that the world trade network is highly centralized in many industries, and notably we show that complex and high-tech goods typically display a stronger centralization of their trade structure. A fundamental result in network science is that the transmission of shocks -- and therefore the vulnerability of the system -- is related to structure, with highly centralized networks being the most fragile \cite{AlJe:00,CoEr:01,BaBa:08,Ba:16}, a feature thoroughly discussed for international trade too \cite{AcCa:12,Ca:14,CoFa:14,KoPi:17}. Given that complex goods are very relevant for all economies, and that high-tech industries -- according to World Bank estimates -- make up about one fifth of all world trade, we argue that the current composition of production is potentially associated to high fragility of the trading system, making it vulnerable to attacks or disasters. The impact of shocks hitting the central nodes in these industries can be large and widespread.

To reach this conclusion, we analyze data of inter-country trade in year 2014 among 223 countries, extracted from the CEPII-BACI database \cite{cepii:17} with HS 4-digit classification, wich defines 1,242 products. We denote by $E=[e_{cp}]$ the 223$\times$1,242 country/product trade matrix, whose entry $e_{cp}$ is the export value (in USD) of product $p$ by country $c$. Several alternative proposals have been put forward to quantify the complexity of a product. To robustify our analysis, we consider three different indicators whose values are computed (or are publicly available) for each one of the products $p$:

\emph{Hidalgo-Hausmann (HH) indicator} ($X'_p$): It is the Product Complexity Index defined in \cite{HiHa:09,HaHi:14}, ranking products by the amount of capabilities or know-how necessary to manufacture them. We downloaded the values of $X'_p$ for the year of interest (2014) from the website of The Atlas of Economic Complexity (http://atlas.cid.harvard.edu/rankings/product/2014/).

\emph{Fitness-Complexity (FC) indicator} ($X''_p$): It is the product complexity measure proposed in \cite{TaCr:12} (extensive metrics form), obtained elaborating on the above HH approach and based on the following non-linear iterative computation:
\begin{eqnarray}\label{fc}
    &\tilde{Q}_p^{(n)}=\frac{1}{\sum_c q_{cp} / F_c^{(n-1)}}, &  \tilde{F}_c^{(n)}=\sum_p q_{cp} Q_p^{(n-1)} ,\\
    &Q_p^{(n)}=\tilde{Q}_p^{(n)}/\langle \tilde{Q}_p^{(n)}\rangle_p,  &  F_c^{(n)}=\tilde{F}_c^{(n)}/\langle \tilde{F}_c^{(n)}\rangle_c   ,
\end{eqnarray}
where $q_{cp}=e_{cp}/\sum_{c'}e_{c'p}$ and $Q_p^{(0)}=1$ $\forall p$, $F_c^{(0)}=1$ $\forall c$. The above iteration is empirically proved to converge \cite{TaCr:12}, and we take the product complexity $X''_p$ as the logarithm of the limit value of $Q_p^{(n)}$.

\emph{PRODY indicator} ($X'''_p$): It is the (weighted) average income per-capita of the countries exporting product $p$ \cite{HaHw:07}:
\begin{equation}\label{prody}
    X'''_p=\sum_c \frac{s_{cp}}{\sum_{c'} s_{c'p}} I_c ,
\end{equation}
where $s_{cp}=e_{cp}/\sum_p e_{cp}$ is the share of product $p$ in the export basket of country $c$, and $I_c$ is the income of country $c$ measured as GDP per capita adjusted for power purchasing parity (data source: The World Bank, https://data.worldbank.org/).

As expected, the three indicators are overall positively correlated (see Supplemental Material \footnote{\label{sm}See Supplemental
Material at [URL will be inserted by publisher] for further statistics on the complexity and centralization indicators.}, Fig. S1), yet they display remarkable differences on many products \cite{HiHa:09,TaCr:12}.

\begin{figure}
\includegraphics[width=8cm]{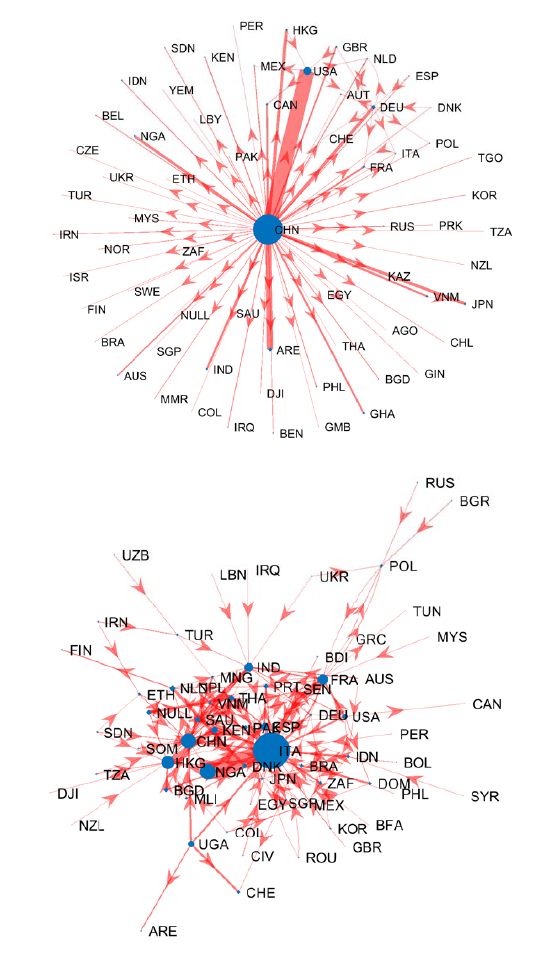}
\caption{\label{f1}Two examples of product trade networks. Above: Product \#8513 ("Portable electric lamps designed to function by their own source of energy", Section "Machinery/Electrical") displays very strong centralization, with most of the export concentrated in one single country and a predominant star-like topology. Below: Product \#4106 ("Tanned or crust hides and skins of other animals, without wool or hair on", Section "Raw Hides, Skins, Leather \& Furs"), on the contrary, is characterized by a few leading countries relating to many others by an intricate pattern of connections. To improve visualization, links carrying less than 1\% of the largest weight have not been displayed.}
\end{figure}

For each product $p$, trade data define a weighted, directed network $N_p$ where the weight $w_{ij}^p$ of the link from country $i$ to country $j$ is the monetary value of the export from $i$ to $j$ (Fig. \ref{f1}). In general terms, a centralization index aims at capturing to what extent a given property is unevenly distributed among network nodes. We are interested, for each product, in quantifying the heterogeneity in the export capabilities of countries. We quantify centralization by three different indicators which not only describe local features (country exports) but also dynamical and robustness properties dependent on the global network structure. All of them take value in the $[0,1]$ range, with zero (resp. one) denoting minimal (resp. maximal) centralization.

\emph{GINI index} ($Y'_p$): A standard Gini index can naturally be used to quantify the unevenness of the distribution of the out-strengths (i.e., the country total export of product $p$). Let $W=[w_{ij}]$ be the weight matrix (we omit the product index $p$ to keep the notation simpler), $s_i^{out}=\sum_j w_{ij}$ the out-strength of node $i$, and $\tilde{W}=[\tilde{w}_{ij}]$ the weight matrix after nodes have been re-ordered by increasing out-strength, i.e., $\tilde{s}_1^{out}\leq\tilde{s}_2^{out}\leq\ldots\leq\tilde{s}_n^{out}$, where $n$ is the number of nodes. We define the cumulated (Lorenz) curve $z_0=0$, $z_i=(1/S)\sum_{j=1}^i \tilde{s}_j^{out}$, $i=1,2,\ldots,n$, where $S=\sum_{ij}w_{ij}$ is the total trade. In the case of least centralization, i.e., all nodes have the same out-strength ($s_i^{out}=S/n$ $\forall i$), the sequence $z_i$ is linearly increasing from 0 to 1. In the opposite case, i.e., only one node has nonzero export while all the others are pure importers, we have $z_0=z_1=\ldots=z_{n-1}=0$, $z_n=1$. The Gini index $Y'_p$ is the normalized distance of the cumulated curve from the linearly increasing curve (see Supplemental Material \footnotemark[1], Fig. S3), so that $Y'_p=0$ for the network with homogeneous out-strength, and $Y'_p=1$ for the star network with the center as the only exporter.

\emph{Core-Periphery (CP) index} ($Y''_p$): It is the core-periphery score defined in \cite{DeDe:13} and based on the notion of persistence probability $\alpha_S$ of a subnetwork $S$, namely the probability that a random walker currently in any of the nodes of $S$ remains in $S$ at the next time step \cite{DeDe:13,Pi:11}. A network profile is built by ordering nodes from the periphery to the core according to a heuristic strategy. We start by the node $i$ with minimal strength, and we generate a sequence $S_1\subseteq S_2\subseteq\ldots\subseteq S_n$ of subnetworks, where $S_1=\{i\}$ is the initial node and $S_n=\{1,2,\ldots,n\}$ is the whole network, by adding at each step $k$ the node attaining the minimal persistence probability $\alpha_k$ of the subnetwork $S_k$. The obtained sequence $0=\alpha_1\leq\alpha_2\leq\ldots\leq\alpha_n=1$ is the core-periphery profile of the network. The complete network and the star network are extreme cases for the core-periphery profile. The former has no core-periphery structure as all nodes are equivalent, so that $\alpha_k=(k-1)/(n-1)$ grows linearly from 0 to 1, while the latter is the most centralized network and has $\alpha_1=\alpha_2=\ldots=\alpha_{n-1}=0$, $\alpha_n=1$. The centralization $Y''_p$ is defined as the normalized distance of the core-periphery profile from that of the complete network (see Supplemental Material \footnotemark[1], Fig. S3), so that $Y''_p=0$ for the complete network and $Y''_p=1$ for the star network.

\emph{Vulnerability index VI} ($Y'''_p$): It is based on \cite{DaBa:06} and measures how rapidly the aggregated network weight is lost when connectivity decreases because nodes are subsequently removed starting from those with largest out-strength. We re-order nodes by decreasing out-strength, thus $\tilde{s}_1^{out}\geq\tilde{s}_2^{out}\geq\ldots\geq\tilde{s}_n^{out}$, and we define the vulnerability profile $1=v_0\geq v_1\geq \ldots \geq v_n=0$, where $v_k$ is the total weight of the network after nodes $\{1,2,\ldots,k\}$ have been removed, divided by the total weight $S$ of the original network. The vulnerability profile falls immediately to zero for a star network ($v_1=0$) whereas it decays linearly for a complete network. We take as $Y'''_p$ the normalized distance of the profile from that of the complete network (see Supplemental Material \footnotemark[1], Fig. S3), so that $Y'''_p=0$ for the complete network and $Y'''_p=1$ for the star network.

\begin{figure}
\includegraphics[width=8.6cm]{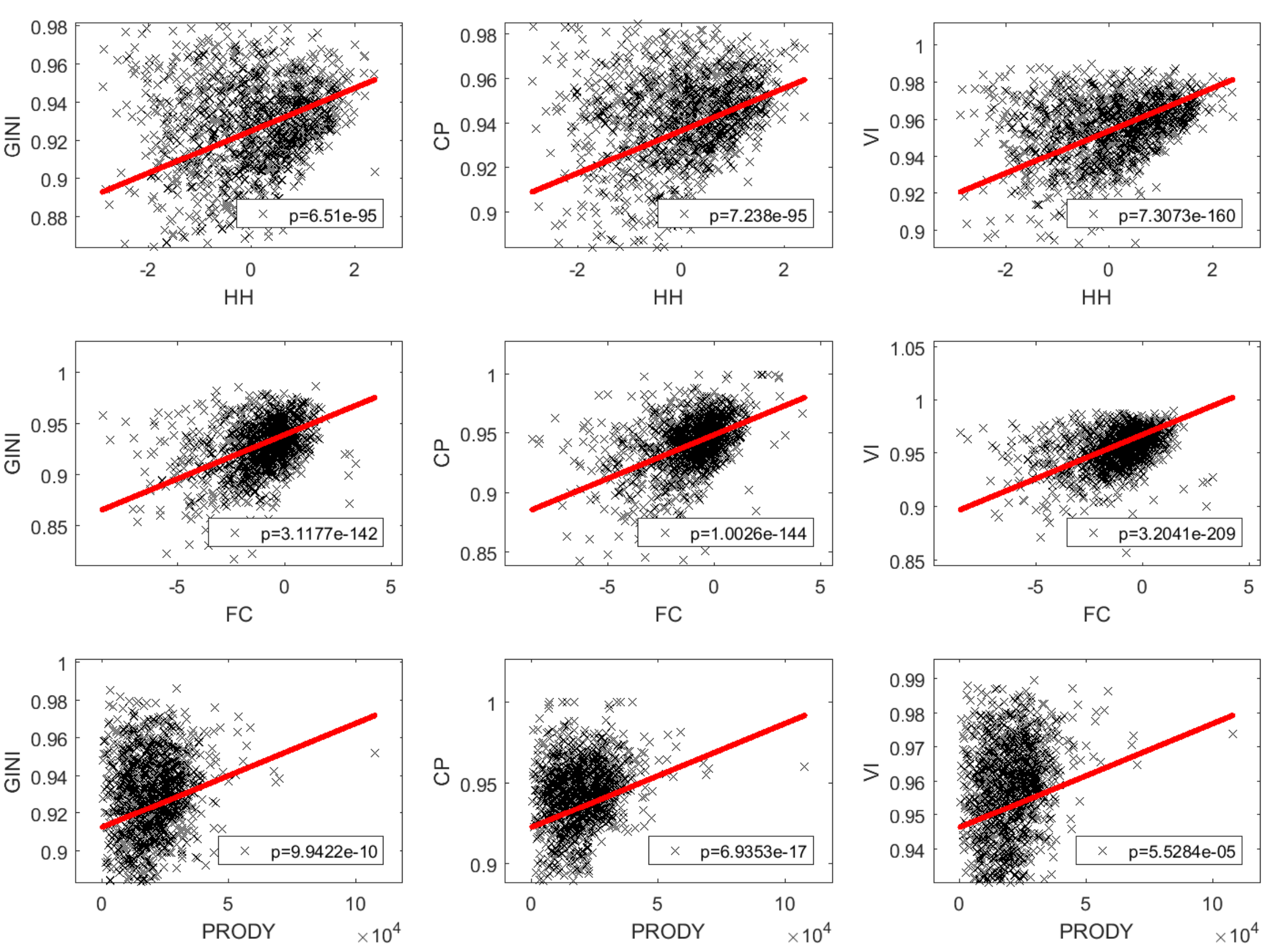}
\caption{\label{f2}Complexity vs centralization in export data. For the $3\times 3$ combinations of indicators, the scatter plots report the complexity/centralization values for the complete set of 1,242 products. The weighted linear regression consistently displays statistically significant positive slope (see p-value in the bottom-right corner).}
\end{figure}

For each one of the $3\times 3$ complexity/centralization pairs $(X_p,Y_p)$ we obtain a scatter plot with one point for each of the 1,242 products $p$. We conjecture that the more complex is a product, the more centralized is its distribution network $N_p$, given that countries with the necessary skills and organization capacity to produce complex products are few compared to countries able to produce and export efficiently simple products. To test this conjecture, we compute the least-squares linear interpolant for each scatter plot. Since the economic importance of products is largely different (see Supplemental Material \footnotemark[1], Fig. S2), we compute a weighted regression, the weight for point $(X_p,Y_p)$ being the total world export $\sum_{ij}w_{ij}^p$ (Fig. \ref{f2}). We expect a positive slope of the linear interpolant, and we check the statistical significance of the result.

Figure \ref{f2} shows that, on the total set of products, complexity and centralization are indeed positively correlated (a restrictive p-value of 1\% is used to check the statistical significance of the positive dependence). Notably, this is consistently true for all the $3\times 3$ complexity/centralization pairs. We corroborated this evidence by an independent analysis where, in place of the above complexity indicators, we use a standardized product classification based on technology content \cite{La:00}: the results (see Supplemental Material \footnotemark[1], Fig. S5) consistently denote an increasing trend of centralization for increasing technological level.

\begin{figure}
\includegraphics[width=7.5cm]{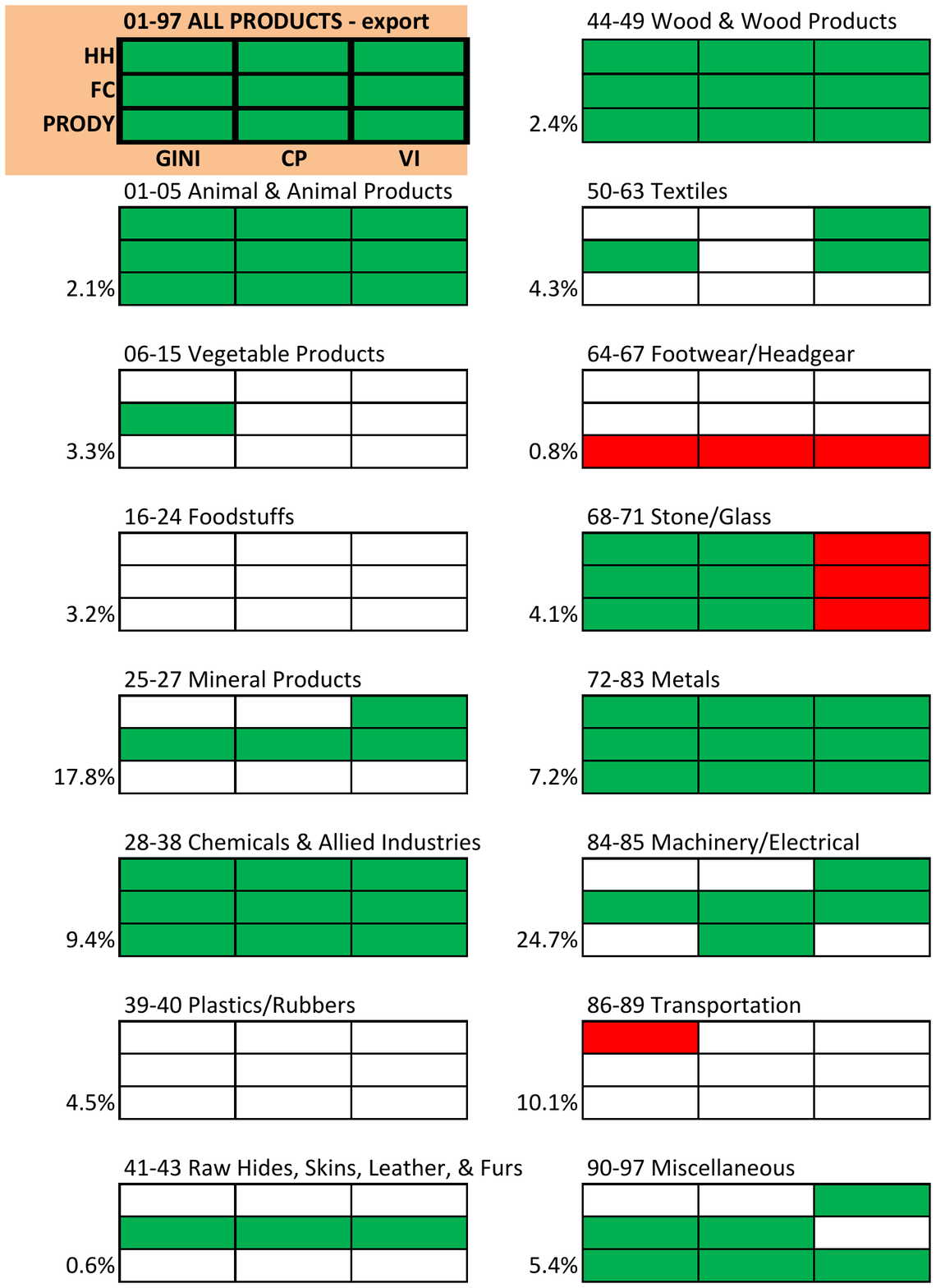}
\caption{\label{f3}Complexity vs centralization in HS Sections export data: consensus analysis. The top-left table refers to the complete set of 1,242 products (Fig. \ref{f2}), the other tables to the specified Section, whose share of the total world trade is on the bottom-left corner. Each table reports the results for the $3\times 3$ combinations of complexity/centralization indicators (see top-left table for details). A green (resp., red) cell denotes a positive (resp., negative) correlation (p-value=1\%); a white cell denotes that correlation is not statistically significant.}
\end{figure}

To discover which categories of products are the main drivers of this pattern, we repeat the same analysis by partitioning the set of products into 15 sets based on the HS Classification by Section \cite{un:02} (Fig. \ref{f3}). Taking into account the relative weight of each Section, i.e., the share of world trade, we have that the Sections most responsible of the overall complexity/centralization pattern are Machinery/Electrical, Chemicals, and Metals. Other Sections have the same consistent behavior (e.g., Animal \& Animal Products) but a rather small trade share, whereas no Section evidences a clear opposite trend.

\begin{figure}
\includegraphics[width=7.5cm]{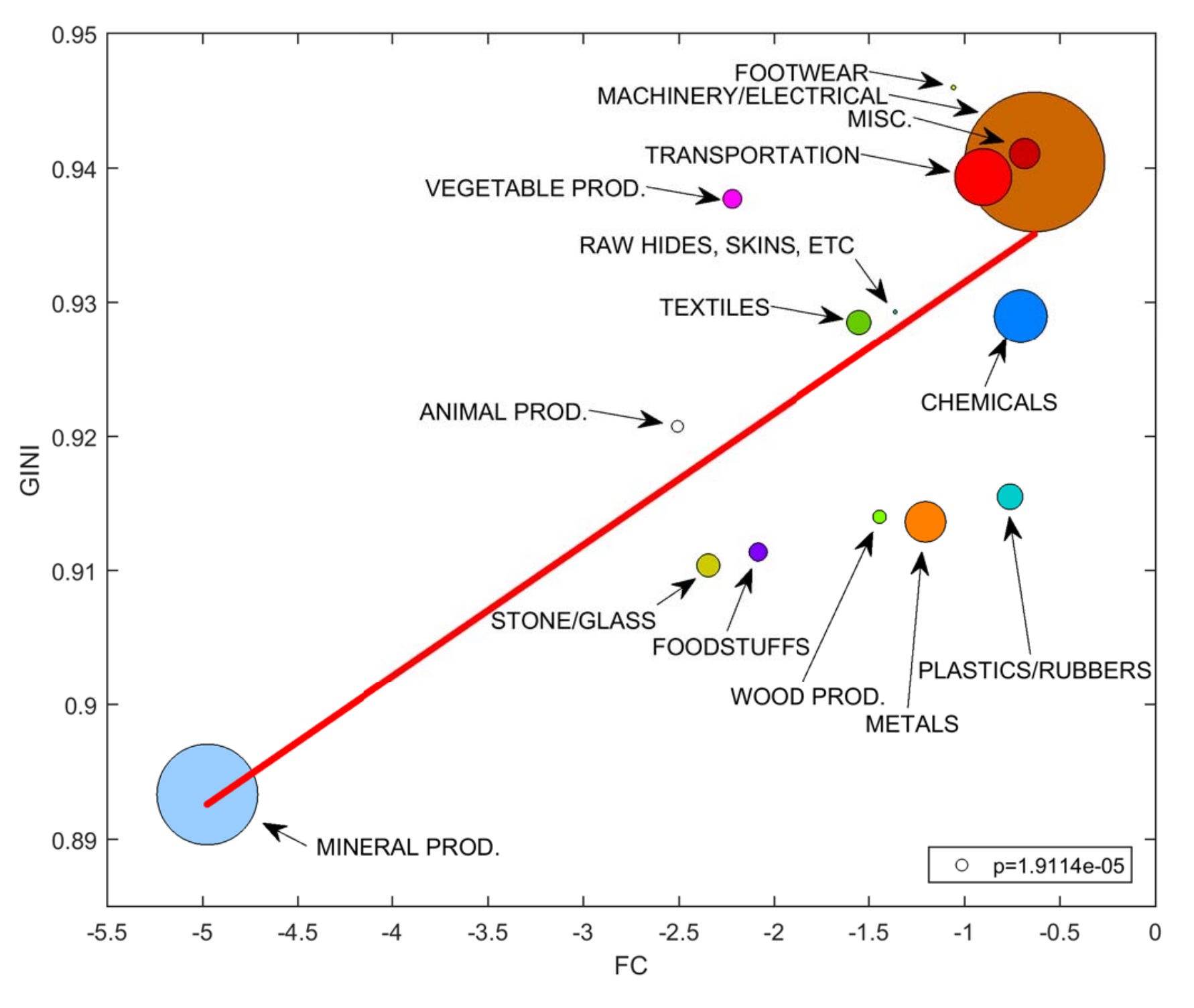}
\caption{\label{f4}Complexity vs centralization in aggregate HS Sections. The scatter plot reports the complexity/centralization values (FC/GINI indicators) for the fifteen Sections. The values are obtained as averages (weighted by trade volume) of the complexities/centralizations of the products of each Section. The marker size is proportional to the total trade volume of the Section. The weighted linear regression (red line) displays statistically significant positive slope (see p-value in the bottom-right corner).}
\end{figure}

Figures \ref{f2} and \ref{f3} show that, typically, products with larger complexity are distributed through a trade network with higher centralization, and that the same holds if we separately consider the most important (in terms of trade volume) subsets of products. The complementary analysis is instead to aggregate products by Section, and to compare the average complexity with the average centralization. The result (Fig. \ref{f4}) confirms that, even at this aggregate level, categories of products with larger complexities are associated to larger centralizations of their trade networks.

The results confirm the conjecture on the positive correlation between complexity of products and centralization of their trade networks. A complex product is obtained by combining different parts and inputs, produced applying specific knowledge and performing particular tasks. These procedures are not easily standardized and their knowledge content not easily transferable, with the possible exception of some limited parts. Therefore, these types of production take place in a small subset of locations and, consequently, complex goods can only be exported by a handful of countries, eventually yielding the observed centralization patterns.

Furthermore, many complex goods are produced through global value chains \cite{Ba:13,ElLo:13}, an organization of production where each phase takes place in a different country to benefit from specific inputs provided more efficiently. This production structure is typically organized around a hub coordinating the whole process. Hence, even if global value chains increase connectivity by generating many trade links between countries exchanging parts and inputs, the complex goods resulting from this organization are eventually exported by the final assembler, giving rise to a centralized structure of trade.  The exceptions to the general correlation pattern refer to groups of products that might not meet the above characterization of complexity for the whole product class, but contain both simple, standardized types of goods, and very complex varieties (e.g., textiles or footwear). Other groups of products displaying a weak correlation are the ones that are not complex but tend to be produced in specific geographical areas (e.g., foodstuffs or wood products) because of the climate or geology of the region, and therefore still tend to have a centralized trade structure.

The high centralization observed for complex products drives the strong hierarchy of the overall trade network, given that they make up an important share of total trade. This implies that the current structure of the world trade network is indeed exposed to specific shocks (e.g., notice that our centralization VI indicator explicitly quantifies the impact of shocks propagation from the central nodes). Considering such structure, it is not surprising that, in 2009, world trade experienced the strongest fall observed for over a century, after a serious economic crisis had hit some of the most central nodes - since then, trade flows have been much more volatile than in the previous decades \cite{BeLe:16}, an undesirable feature that, given the persistence of such structure, could continue for long.  While the literature highlights that uncertainty at the country level can have detrimental effects on local trade \cite{AnMa:02}, fragility can play a similar effect on trade at the global level.

\begin{acknowledgments}
We thank E. Marvasi for assistance with data retrieval and pre-processing.
\end{acknowledgments}

%%%%%%%%%%%%%%%%%%%%%%%%%%%%

%\bibliography{../../jlist_short,../../my_biblio}


%%%%%%%%%%%%%%%%%%%%%%%%%%%%%

\newpage
\begin{figure*}
\includegraphics[width=18cm]{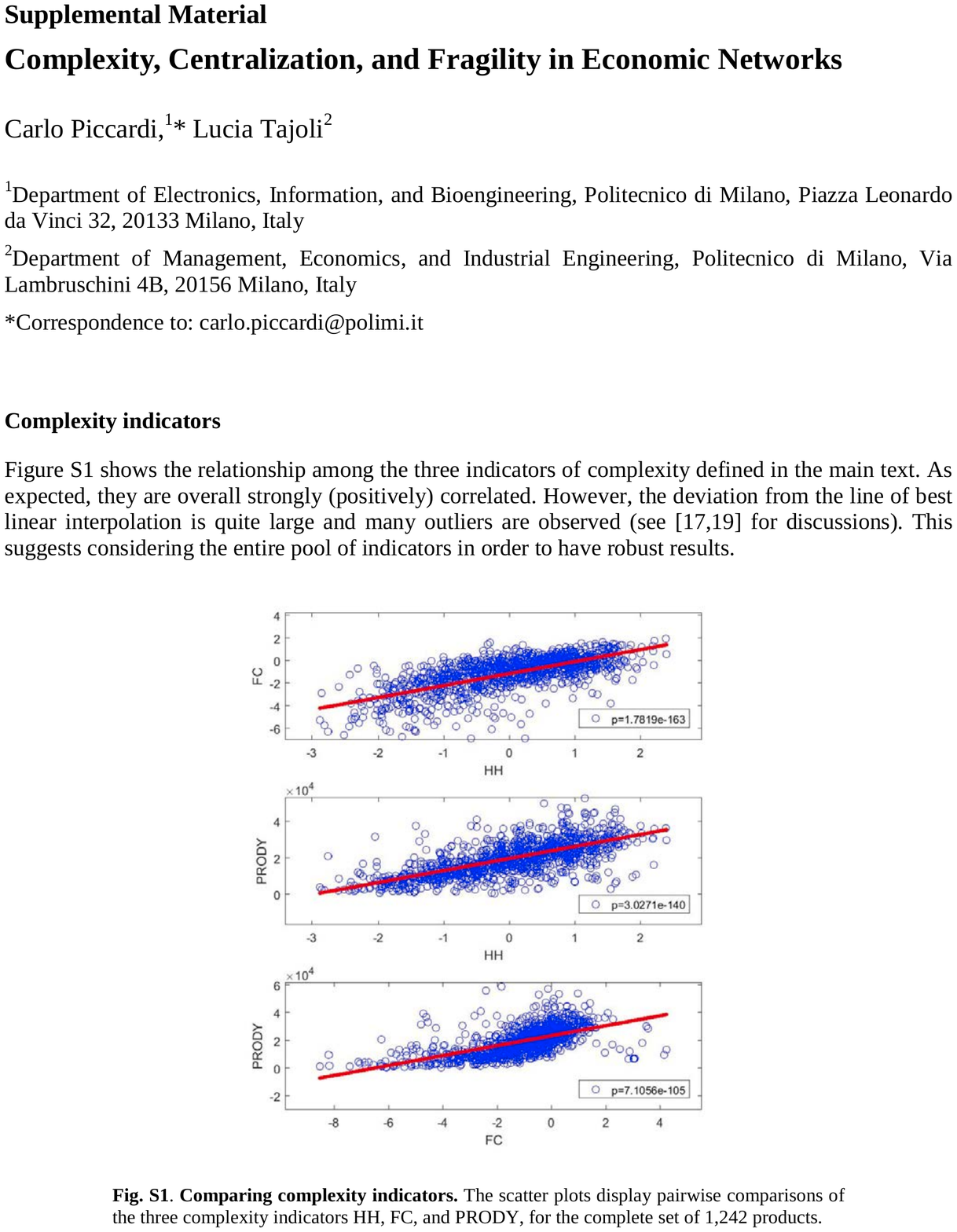}
\end{figure*}
\newpage
\begin{figure*}
\includegraphics[width=18cm]{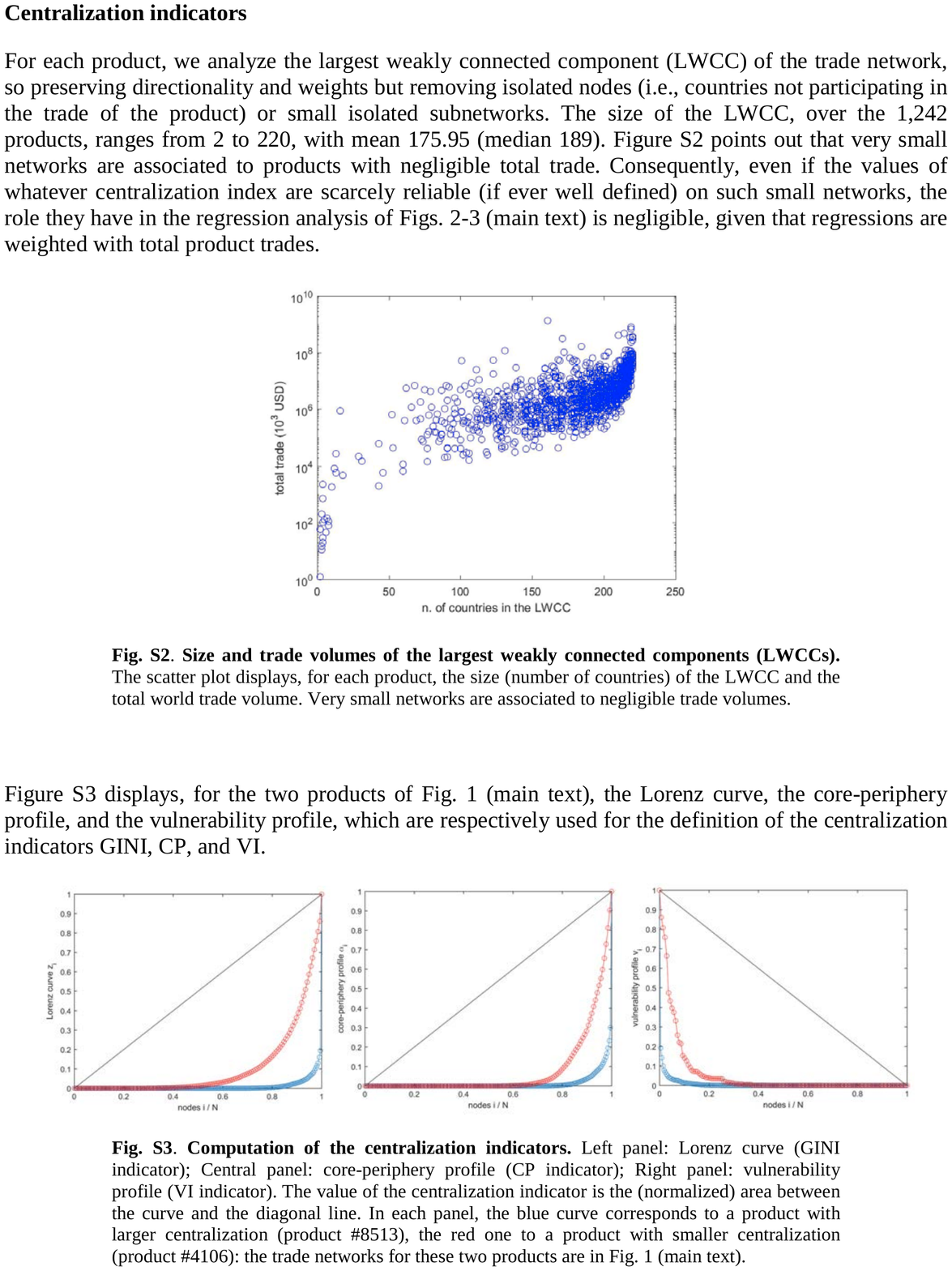}
\end{figure*}
\newpage
\begin{figure*}
\includegraphics[width=18cm]{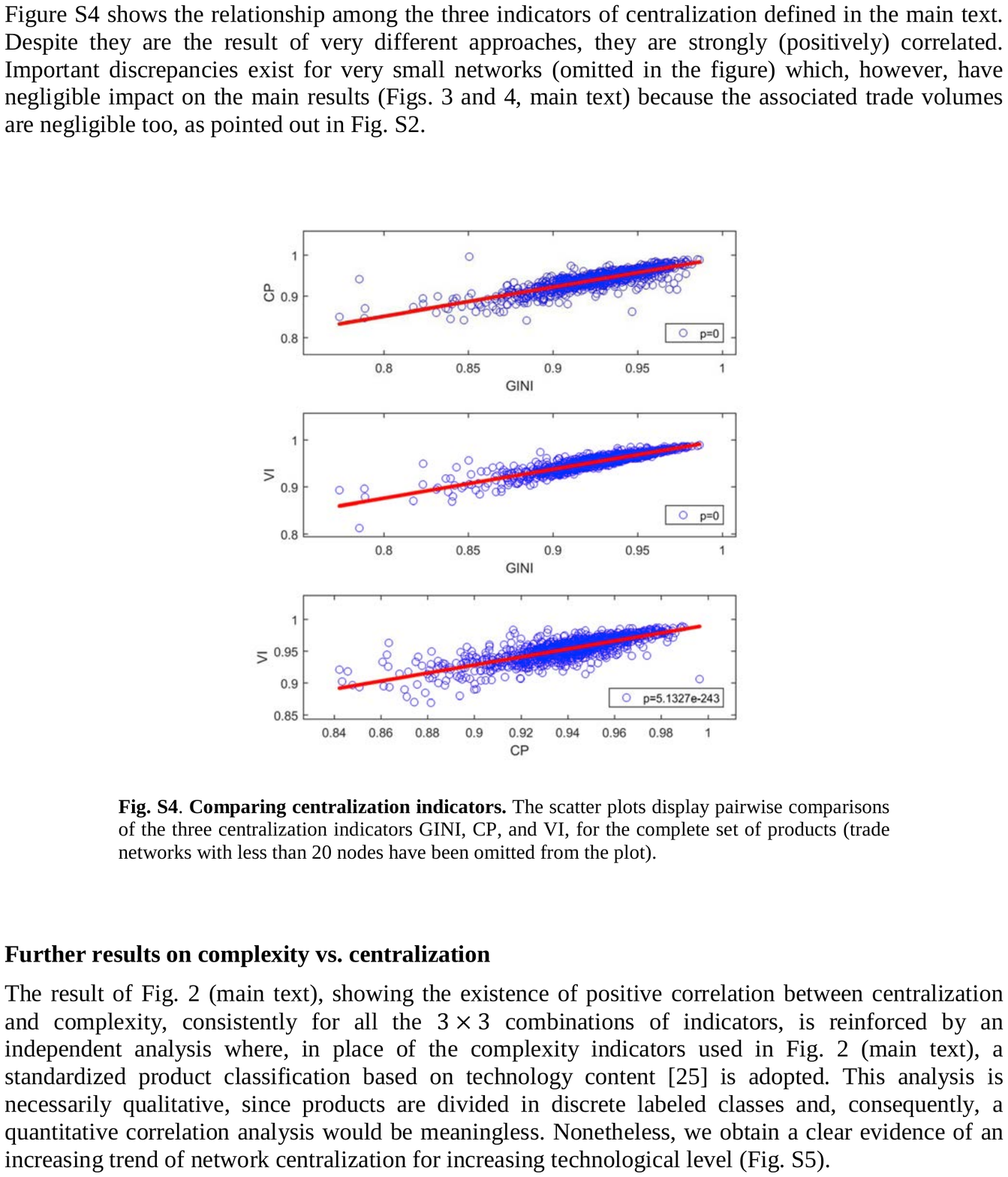}
\end{figure*}
\newpage
\begin{figure*}
\includegraphics[width=18cm]{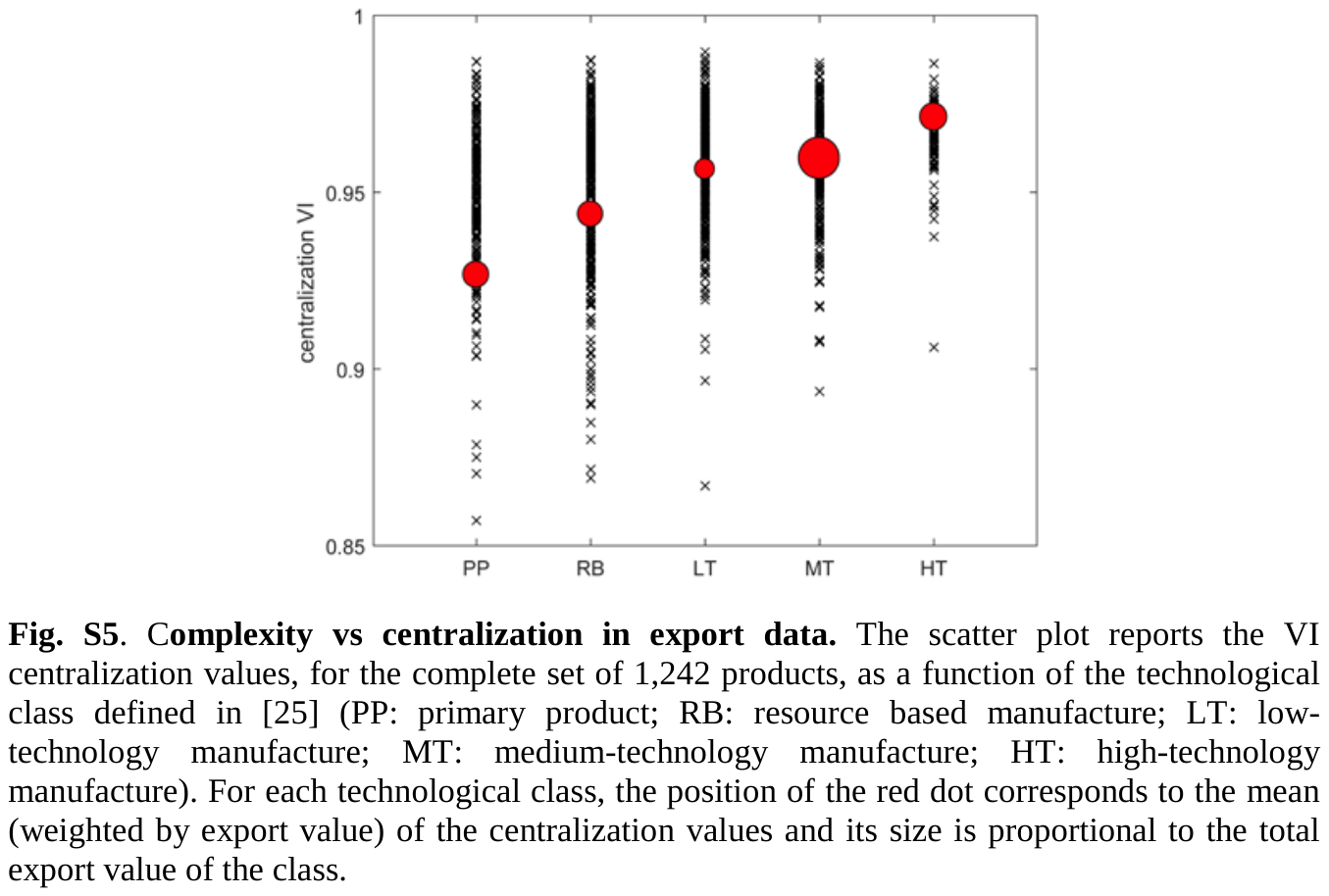}
\end{figure*}

\end{document}